\documentclass[modern]{aastex631}
\usepackage{amsmath}

\newcommand{\pdf}{\mathrm{Pr}}

\shorttitle{Princess Bride Syndrome II}
\shortauthors{Kipping \& Benneke}
\graphicspath{{./}{figures/}}

\begin{document}

\title{Exoplaneteers Keep Overestimating Sigma Significances}

\author[0000-0002-4365-7366]{David Kipping}
\affiliation{Dept of Astronomy, Columbia University, 550 W 120th Street, New York, NY 10027}
\author[0000-0001-5578-1498]{Bj\"orn Benneke}
\affiliation{Dept of Earth, Planetary, and Space Sciences, University of California, Los Angeles, CA USA}
\affiliation{Department of Physics and Trottier Institute for Research on Exoplanets, Université de Montréal, Montreal, QC, Canada}



\begin{abstract}
Astronomers, and in particular exoplaneteers, have a curious habit of expressing Bayes factors as frequentist sigma values. This is of course completely unnecessary and arguably rather ill-advised. Regardless, the practice is common - especially in the detection claims of chemical species within exoplanet atmospheres. The current canonical conversion strategy stems from a statistics paper from \citet{sellke:2001}, who derived an upper bound on the Bayes factor between the test and null hypotheses, as a function of the $p$-value (or number of sigmas, $n_{\sigma}$). A common practice within the exoplanet atmosphere community is to numerically invert this formula, going from a Bayes factor to $n_\sigma$. This goes back to \citet{benn:2013} --- a highly cited paper that introduced Bayesian model comparison as a means of inferring the presence of specific chemical species --- in an attempt to calibrate the Bayes factors from their technique for a community that in 2013 was more familiar with frequentist sigma significances. However, as originally noted by \citet{sellke:2001}, the conversion only provides an upper limit on $n_\sigma$, with the true value generally being lower. This can result in inflations of claimed detection significances, and this note strongly urges the community to stop converting to $n_\sigma$ at all and simply stick with Bayes factors.


%
\end{abstract}




\section*{}

\begin{figure}[t]
\begin{center}
\includegraphics[width=12.0cm,angle=0,clip=true]{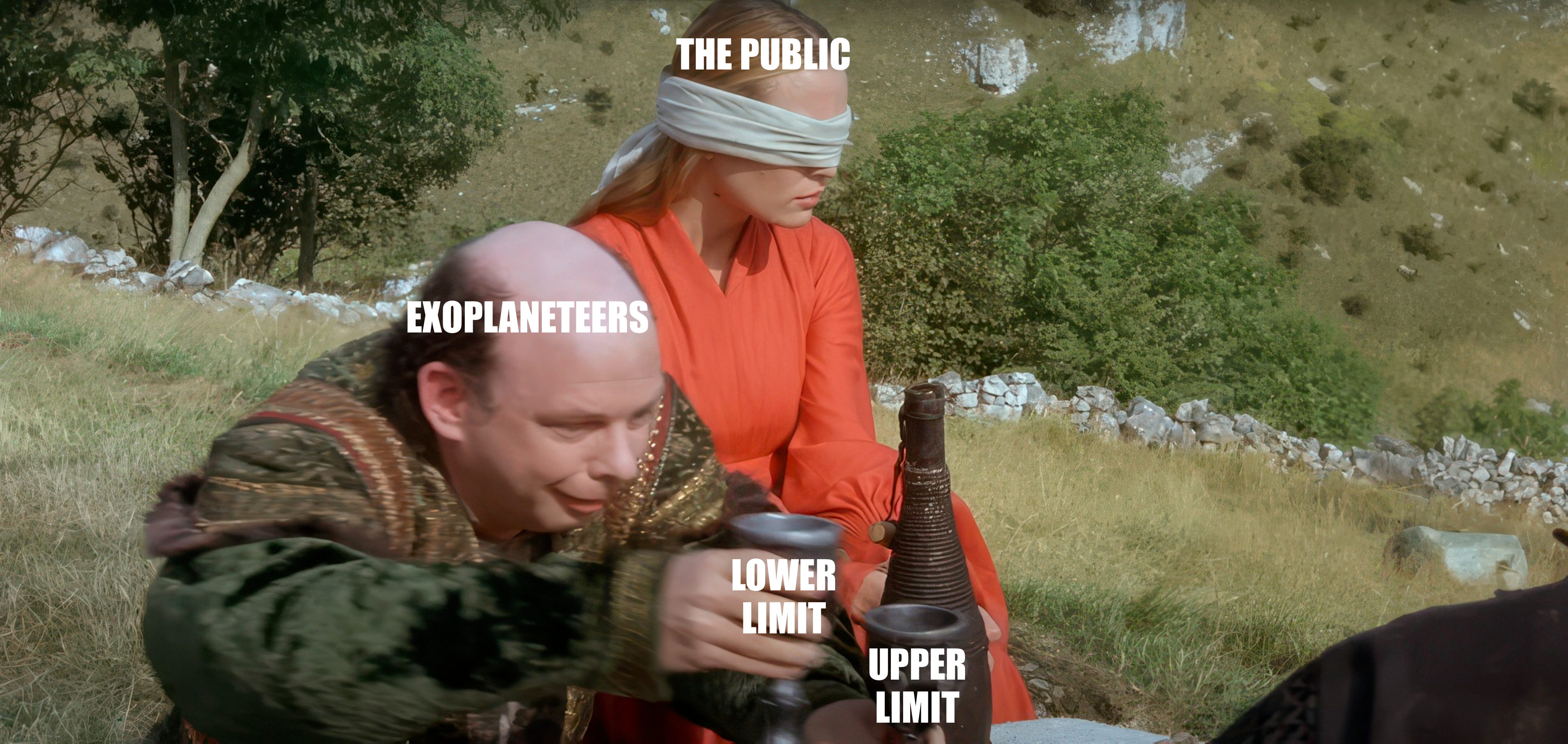}
\caption{\textit{
``What in the world could that be?'' Vizzini, The Princess Bride.
}}
\label{fig:vizzini}
\end{center}
\end{figure}

\section{Converting Bayes Factors into Sigmas}

At the time of writing, the use of Bayesian inference techniques is widespread amongst astronomers \citep{eadie:2023}. Bayesian model selection has emerged as the canonical tool when seeking to detect some phenomenon of interest, such as the spectral absorption feature of a particular chemical species. To the Bayesian, the data ($\mathcal{D}$) are fixed but the hypotheses (and model parameters) are probabilistic and thus all one can do is rank hypotheses against one another, most commonly achieved using odds ratios e.g. $\pdf(\mathcal{H}_1|\mathcal{D})/\pdf(\mathcal{H}_0|\mathcal{D})$. So, for example, hypothesis $\mathcal{H}_1$ may represent the inclusion of some phenomenon into a broader model, and $\mathcal{H}_0$ represents the ``vanilla'' broader model (i.e. the null hypothesis) which does not include it. In this way, detections can often be framed as the act of Bayesian model selection between \textit{nested} hypotheses.

This odds ratio equals the Bayes factor ($\pdf(\mathcal{D}|\mathcal{H}_1)/\pdf(\mathcal{D}|\mathcal{H}_0)$) multiplied by the hypotheses' prior ratio ($\pdf(\mathcal{H}_1)/\pdf(\mathcal{H}_0)$) - which is typically assumed to be unity i.e. agnostic. Thus, the Bayes factor dominates discussions of detection significance. A Bayes factor of $X$ can be interpreted as the following: ``The data are $X$ times more likely under model 1 than under model 0''. That's really about all we can say and strictly speaking there is no magical threshold at which point $X$ becomes a ``detection''.

Of course, this presents a challenge to scientists presenting their work to the public and even the broader community. Bayes factors are subtle and unfamiliar to those not versed in statistical inference. One approach is to neatly classify Bayes factors into buckets, such as the Jeffrey's scale \citep{jeffreys:1939} or that of \citet{kass:1995}. Another more precarious strategy is to attempt to convert Bayes factors into ``sigmas'', presumably because there is a perception that sigmas are more familiar conceptually. It's possible this perception became popularized by the sensational detection of the Higgs boson at the $5$\,$\sigma$ level \citep{higgs:2012}, which amplified the notion of $5$\,$\sigma$ as the gold-standard for unambiguous discoveries\footnote{Of course, nothing magical happens from $4.9$ to $5.0$\,$\sigma$.}. Regardless, the conversion is problematic as one is attempting to graft the Bayesian worldview onto that of the frequentist \citep{trotta:2008}.

\section{The Sellke et al. Formula}

\citet{sellke:2001} derived a formula for this correspondence under a set of basic assumptions: i) the null hypothesis is assumed to be a ``precise'' hypothesis e.g. $\mathcal{H}_1$: $\theta=0$; ii) the alternative is a composite hypothesis thereby including range of values e.g. $\mathcal{H}_2$: $\theta\neq0$; iii) the problem is univariate; iv) the likelihood ratio is monotonic and continuous; v) the prior is arbitrary but proper; and, vi) the marginal likelihood is well-defined (i.e. finite). Under these assumptions, \citet{sellke:2001} obtain, in their Equation~(2):

\begin{align}
B_{01} &\geq -\exp p \log p,
\label{eqn:sellke}
\end{align}

where $B_{01}$ is the Bayes factor of model 0 to model 1 ($\pdf(\mathcal{D}|\mathcal{H}_0)/\pdf(\mathcal{D}|\mathcal{H}_1)$) and $p$ is the $p$-value of obtaining the data under model 0 (the null). We have made two minor changes in Equation~(\ref{eqn:sellke}) to that of Equation~(2) of \citet{sellke:2001}. First, \citet{sellke:2001} use a ``$=$'' sign rather than a ``$\geq$'' sign, but clearly state after the formula that they ``interpret this as a lower bound on the odds provided by the data (or Bayes factor) for $\mathcal{H}_0$ to $\mathcal{H}_1$''. Second, again based on that quote, we wrote $B_{01}$ as the subject to denote the direction of the odds ratio, whereas \citet{sellke:2001} originally simply wrote $B$.

It's worth briefly considering an example to see what this formula is really saying. And fortunately \citet{sellke:2001} give one: ``Thus, $p=0.05$ translates into odds $B=0.407$ (roughly 1 to 2.5) of $\mathcal{H}_0$ to $\mathcal{H}_1$''. They then go on to write that ``Clearly $p=0.05$ does not indicate particularly strong evidence against $\mathcal{H}_0$''. This example captures the spirit of their underlying argument - that there is a widespread fallacy that a $p$-value such as $0.05$ implies compelling evidence, whereas \citet{sellke:2001} argue that the corresponding Bayes factor can be very modest. 

To our knowledge, the first time the \citet{sellke:2001} formula was first introduced to the astronomy community occurs in Section~4.5 of the classic Bayesian primer of \citet{trotta:2008}. In Equation~(27) of that work, \citet{trotta:2008} flips the odds ratio to the more conventionally stated ratio of the test hypothesis against the null:

\begin{align}
B_{10} &\leq \bar{B_{10}} = -\frac{1}{\exp p \log p},
\label{eqn:trotta}
\end{align}

where $\bar{B_{10}}$ is the upper limit on $B_{10}$. Note that the inequality direction has reversed in this expression (versus that of Equation~\ref{eqn:sellke}) as a result of the flip. It's also worth noting that there appears to be no mention in Section~4.5 of \citet{trotta:2008} of the notion of inverting the \citet{sellke:2001} formula to solve for $p$, given some input $B_{10}$. That concept is discussed, though, in a highly influential exoplanet atmospheres paper by \citet{benn:2013} - although this may not be the first ever such instance of someone attempting this.

The paper by \citet{benn:2013} is primarily focused on introducing a Bayesian framework for detecting chemical species, advocating for an explicit leave-one-out methodology of computing the Bayesian factors between one retrieval model that should cover the full prior hypothesis space and retrieval model for which selectively one individual molecular species (or type of aerosol) was removed from that otherwise full prior hypothesis space. 

However, as a minor note in this paper, \citet{benn:2013} also provided the backward conversion from Bayes factors to sigmas in an attempt to calibrate the Bayes factors in response to members of the community being so unfamiliar with Bayesian model comparison that they were uncomfortable interpreting Bayes factors as a measure of how convincing a particular detection is. Whilst never intended to be broadly used in this way, the community subsequently latched onto this conversion and it has become a widespread practice that often loses sight of the original source. As a recent example (amongst many), \citet{radica:2025} perform this conversion even referring to it as the ``\citet{benn:2013} scale'', presumably unaware of the original \citet{sellke:2001} paper.

This calibration to sigma values is problematic. To see why this, we start with Equation~(10) of \citet{benn:2013}, which (under the assumptions made in \citealt{sellke:2001}) correctly stated

\begin{align}
B_{10} &\leq -\frac{1}{\exp p \log p}.
\label{eqn:benn}
\end{align}

After this equation, \citet{benn:2013} provided the conversion from $p$ to $n_{\sigma}$ (number of sigmas), which we write here as $p = \mathrm{erfc}[n_{\sigma}/\sqrt{2}]$. Unfortunately, however, the fact that Equation~(10) of \citet{benn:2013} has an $\leq$ sign and not an $=$ sign has too often been ignored in the subsequent literature, and a typographical error in one explanatory sentence in text of \citet{benn:2013} itself may have added to the confusion. \citet{benn:2013} correctly stated that ``Equation (10) presents an upper bound on the Bayes factor''; however, that means that a Bayes factor of for examples $B_{10} = 21$ corresponds, at most, to a 3.0\,$\sigma$, and \textit{not} ``at least a 3.0\,$\sigma$'' - as appeared in this paper.

To illustrate this, consider just the first part of the statement: ``Equation (10) presents an upper bound on the Bayes factor''; in this case that's 21. The true Bayes factor could therefore be lower - say, 15. Taking this value of 15, inverting Equation~(\ref{eqn:benn}) yields a $p$-value of 0.00454..., or approximately 2.8\,$\sigma$. Thus, a Bayes factor of 21 does not necessarily correspond to at least a 3.0\,$\sigma$ detection, as it is also consistent with 2.8\,$\sigma$, or indeed values even less than this.

In summary, there is nothing intrinsically wrong with the \citet{sellke:2001} formula for relating Bayes factors and sigmas. But, if one uses it to convert a Bayes factor into $n_{\sigma}$, it must be understood that the sigma value returned is the most optimistic interpretation of how significant the detection truly is, and the true number of sigmas will - in general - \textit{be less}. Indeed, the original use of the upper limit on $B_{10}$ was to discount the possibility of a detection when the limit is not large, since there is no other reference prior that can yield a higher probability e.g. see \citet{gordon:2007}.

There is a certain irony that \citet{sellke:2001} were trying to argue that if one takes typical sigma scores and convert them into the most conservative possible Bayes factor, the odd ratios can be quite modest. In other words, scientists were often overestimating their confidence. It would seem the formula was never really intended to be used the other way round - to convert Bayes factors into sigmas - since that clearly returns the most optimistic possible sigma score, which is of questionable utility and certainly goes against the spirit of Sellke's argument: a plea for conservatism.

\section{$\sigma$-Inflation}

The danger of using the formula is that relatively modest Bayes factors can be converted into surprisingly large sigma values. For example, a Bayes factor of 3 yields 2\,$\sigma$. This can be misleading, as a 3:1 odds factor might naturally suggest a 25\% false-positive rate, whereas a 2\,$\sigma$ significance is often associated with only a 5\% rate. Of course, the reason is that this is merely the absolute maximum possible sigma score possible, and the true value will be lower. There is, then, a danger in authors calculating Bayes factors and converting them into sigmas using the \citet{sellke:2001} formula, without appreciating that this is a highly optimistic and inflated value.

Equation~(10) of \citet{benn:2013} illustrates a common phenomenon: the widespread adoption of a result derived elsewhere, which gains prominence through its contextual use rather than original derivation. The result itself was not derived in that paper, but rather in \citet{sellke:2001}; however, \citet{benn:2013} introduced it to the exoplanet community for the first time. But the frequent lack of original source citation within the field suggests that many researchers may be relying on secondary sources, such as \citet{benn:2013}, rather than consulting \citet{sellke:2001} directly.

A particularly notable example is the recent claim of 3\,$\sigma$ evidence for DMS/DMDS in the atmosphere of K2-18\,b \citep{madhu:2025}. We cite this example here purely as a prominent recent example of a widespread practice, and not as a critique of the authors' intent or work. Their Table~2 provides both the Bayes factors and $n_{\sigma}$ conversions and thus we confirmed these are precisely the values one would obtain using the formula of \citet{sellke:2001}. Despite this, neither \citet{sellke:2001} nor \citet{benn:2013} are cited by \citet{madhu:2025} making it challenging to assess the broader prevalence of this issue via ADS citation tracking. We highlight that this lack of primary source citation is reminiscent of the issue described in a previous commentary about the Allan variance \citep{allan:2025}. From Table~2 of \citet{madhu:2025}, the Bayes factors range from 17.5 to 68.0, and that lowest value corresponds to 2.9\,$\sigma$ using the \citet{sellke:2001} formula. Accordingly, the abstract of \citet{madhu:2025} stated ``We report new independent evidence for DMS and/or DMDS in the atmosphere at 3-$\sigma$ significance'' - whereas truthfully this should be rephrased to ``We report new independent evidence for DMS and/or DMDS in the atmosphere at \textbf{less than} 3-$\sigma$ significance'', in order to match the direction of the \citet{sellke:2001} inequality. This problem is then exacerbated by the \href{https://www.cam.ac.uk/stories/strongest-hints-of-biological-activity}{press release} issued by Cambridge University, which stated ``The observations have reached the `three-sigma' level of statistical significance – meaning there is a 0.3\% probability that they occurred by chance'', whereas again the significance is likely overestimated following the direction of the inequality of \citet{sellke:2001}. In our opinion, it is far better to simply state the Bayes factor - 17:1.

\section{What Should We Do, Then?}

\begin{figure}[b]
\begin{center}
\includegraphics[width=12.0cm,angle=0,clip=true]{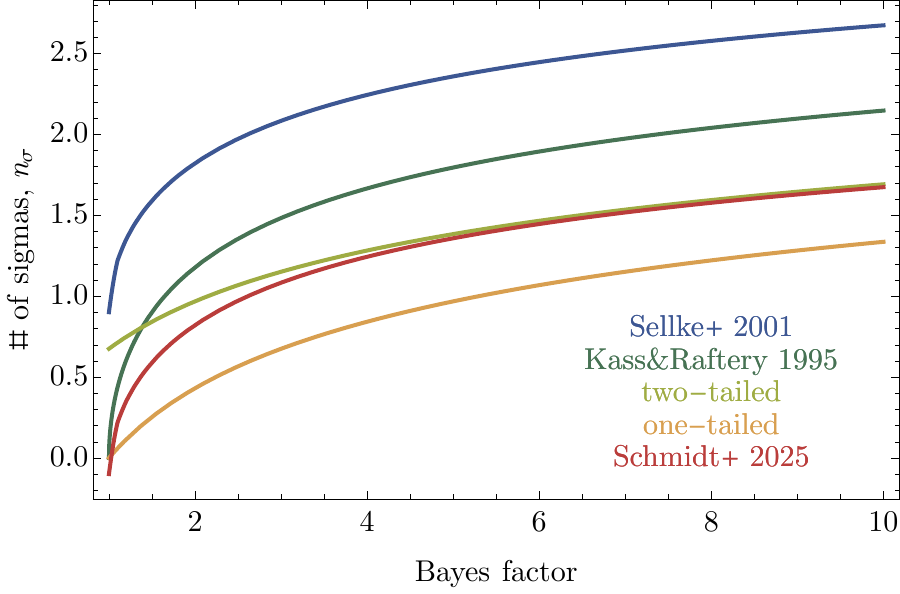}
\caption{\textit{
Five schemes for converting Bayes factors into sigmas. The \citet{sellke:2001} scheme produces the most optimistic values and should be understood as the ceiling.
}}
\label{fig:2}
\end{center}
\end{figure}

Other schemes exist for converting Bayes factors into sigmas. Perhaps the most intuitive is to argue that a $B$:1 odds implies a $p$-value of $1/(B+1)$, which follows from a two-tailed $p$-value and assumes only two hypotheses exist. This scheme is reasonable and certainly more conservative than inverting the formula of \citet{sellke:2001}, as Figure~\ref{fig:2} illustrates. However, it comes with an offset problem: a Bayes factor of 1 implies a $p$-value of 50\%, which converts to $0.7$\,$\sigma$. Of course, a Bayes factor of 1 means there is no evidence whatsoever for the hypothesis, but even here someone ignorant of this nuance could argue they have a weak $\simeq 1$\,$\sigma$ claim. If there is a widespread intuition to interpret sigmas as some kind of confidence score (however misguided that may be; see \citealt{hubbard:2008}), then one should expect a Bayes factor of 1 to return $n_{\sigma}=0$.

As a compromise, \citet{schmidt:2025} argue for taking the \citet{sellke:2001} formula but subtracting one off the resulting number of sigmas\footnote{I also note that \citet{trotta:2008} allude to this idea in their Section~4.5.} - this produces a conservative conversion which asymptotically approaches the two-tailed formula, but returns $-0.1$\,$\sigma$ for $B=1$ and lacks a rigorous underpinning. In private correspondence, Michael Zhang suggested a one-tailed $p$-value provides an alternative means of fixing the offset problem, such that $p = 2/(B+1)$ (e.g. $B=1$ yields $p=1$ and $n_{\sigma}=0$). I show this scheme in Figure~\ref{fig:2}, which produces the most conservative scheme.

An alternative formalism is that of \citet{kass:1995}, who propose $n_{\sigma} \simeq \sqrt{2 \log B}$, valid in the case of nested models (which is generally true) and a large number of data points (not necessarily true e.g. binned spectra). This has the desirable property of tending to zero as $B\to1$ and returns values in between the two-tailed scheme and that of \citet{sellke:2001} - see Figure~\ref{fig:2}. A comparison of the five schemes is presented in Figure~\ref{fig:2}.

None of these schemes are ideal and arguably the entire exercise is ill-advised and unnecessary. We suggest it is better to simply stick to Bayes factors. Concerning public communication, we would further argue that odds ratios are more intuitive than sigmas anyway due to their association with gambling and risk assessment, and our job as communicators should be to explain the nuance where present.

\vspace{2cm}
Thanks to Roberto Trotta, Ryan Macdonald, Daniel Yahalomi and Ben Cassese for useful conversations in preparing this note.
Special thanks to Michael Zhang for his suggestion regatrding the one-tailed $p$-value.

%

\vspace{5mm}






\bibliography{manuscript}{}
\bibliographystyle{aasjournal}



\end{document}